\documentclass[prb,aps,twocolumn,floatfix]{revtex4}
\usepackage{amsmath,graphics,epsfig,color,verbatim,ulem}

\begin{document}

\title{A dynamical mean-field theory study of Nagaoka ferromagnetism.}

\author{Hyowon Park, K. Haule, C. A. Marianetti,  and G. Kotliar }

\date{\today}

\begin{abstract}
We revisit Nagaoka ferromagnetism in the $U=\infty$ Hubbard model
within the dynamical mean-field theory (DMFT) using the recently developed
continuous time quantum Monte Carlo method as the impurity solver.
The stability of Nagaoka ferromagnetism is studied as a function of
the temperature, the doping level, and
the next-nearest-neighbor lattice hopping $t^{\prime}$.
We found that the nature of the phase transition as well as the
stability of the ferromagnetic state is very sensitive to the
$t^\prime$ hopping. Negative $t^{\prime}=-0.1t$ stabilizes
ferromagnetism up to higher doping levels. The paramagnetic state is
reached through a first order phase transition.  Alternatively, a
second order phase transition is observed at $t^\prime=0$.
Very near half-filling, the coherence temperature $T_{coh}$ of the
paramagnetic metal becomes very low and ferromagnetism evolves out of
an incoherent metal rather than conventional Fermi liquid.
We use the DMFT results to benchmark slave-boson method which might be
useful in more complicated geometries.
\end{abstract}

\affiliation{Department of Physics and Astronomy and Center for Condensed Matter
Theory, Rutgers University, Piscataway, NJ 08854--8019}

\maketitle

\section{Introduction}

The stability of the ferromagnetic phase in the $U=\infty$ Hubbard
model is a long standing problem.  Nagaoka~\cite{Nagaoka:66} showed
that for a single hole in a bipartite lattice the ground state is a
fully polarized ferromagnet, and the term "Nagaoka ferromagnetism" is
commonly used to describe this state.  Whether a fully or a partially
polarized phase persist to a finite hole density ($\delta$) is
controversial and has been the subject of numerous
investigations~\cite{Patrick}.

The problem has been addressed with variational wave functions
~\cite{Shastry:90,Basile:90,Linden:91,Hanisch:93,Wurth:96}, slave
particle methods ~\cite{Moller:93,Boies:95}, quantum Monte Carlo (QMC)
methods~\cite{Zhang:91}, and variational QMC methods~\cite{Becca:06}.
In all these methods the ferromagnetism is stable up to a critical
value of doping $\delta_{c}$. It was also demonstrated by these
approaches that the size of the ferromagnetic region depends strongly
on the lattice through the electronic dispersion.
The ferromagnetic state was found to be unstable even for the case of
a single hole in the $U=\infty$ square lattice with a small positive
next-nearest neighbor hopping $t^{\prime}$~\cite{Oles:91}.
At an intermediate or a large $U$, a flat band below the Fermi level
~\cite{Tasaki:92} or a peak in the density of states below the Fermi
level ~\cite{Gagliano:90,Hanisch:97,Wahle:98,Arrachea:00}, as realized
in the fcc lattice~\cite{Ulmke:98,Wegner:98} or a Van Hove
singularity~\cite{Hlubina:97}, stabilize the ferromagnetic state.

The dynamical mean-field theory (DMFT) has also been used to address
the Nagaoka problem, however the number of available impurity solvers
in the $U=\infty$ case is very limited.  Obermeier \textit{et
al.}~\cite{Obermeier:97} carried out the first DMFT study of this
problem using the non-crossing approximation as the impurity solver.
They found a partially polarized ferromagnetic state below a critical
temperature $T_{c}$ in the infinite dimensional hypercubic lattice.
The existence of a ferromagnetic state in this model was
later confirmed by a DMFT study which used numerical renormalization group as the impurity
solver~\cite{Zitzler:02}.

In this study, we revisit the problem of Nagaoka ferromagnetism in the
$U=\infty$ Hubbard model within DMFT, using the recently developed
continuous time quantum Monte Carlo (CTQMC) method as the impurity
solver \cite{Haule:06,Werner:06}. This impurity solver allows the
numerically exact solution of the DMFT equations at very low
temperatures for all values of doping level $\delta$ even in the
$U=\infty$ model.  We find that at large doping, the ferromagnetism
emerges from a conventional Fermi liquid, while at small doping the
Curie temperature is very close to the coherence temperature, hence
the ferromagnetism emerges from an incoherent state.  We pay
particular attention to the possibility of phase separation and its
dependence on the sign of $t^{\prime}/t$.  Finally we benchmark
simpler approaches to the problem such as the slave boson method.
Within slave boson approach, several physical quantities such as the
quasiparticle renormalization amplitude or the susceptibility can not be
determined reliably.  Nevertheless we show that the total energy
can be computed quite reliably within the simple slave boson approach
due to error cancellation.  This is important since the detailed
modeling of optical lattices of cold atoms, which provide a clean
realization of the Hubbard model, will require incorporating spatial
inhomogeneities into the treatments of strong correlations.  At
present, this can only be done with simpler techniques such as slave
bosons methods.

We study the Hamiltonian of the $U=\infty$ Hubbard model given by
\begin{eqnarray}
   \hat{H}=-\sum_{ij\sigma}t_{ij}\hat{P_{s}}\hat{c}^{\dag}_{i\sigma}\hat{c}_{j\sigma}\hat{P_{s}},
\end{eqnarray}
where $\hat{P_{s}}$ is a projection operator which removes states with
double-occupied sites.  We choose the lattice dispersion of the two
dimensional square lattice with the nearest-neighbor (n.n) hopping $t$
and the next-nearest-neighbor (n.n.n) hopping $t^{\prime}$.
The units
are fixed by choosing $t=\frac{1}{2}$.

%
%


\section{A DMFT+CTQMC approach}

DMFT maps the partition function of the Hubbard model onto the
partition function of an effective Anderson impurity model (AIM)
resulting in the following effective action.

\begin{eqnarray} \label{action}
   S_{eff}=S_{atom}+\int_{0}^{\beta}d\tau\int_{0}^{\beta}d\tau^{\prime}\sum_{\sigma}c_{\sigma}^{\dagger}(\tau)\Delta_{\sigma}(\tau-\tau^{\prime})c_{\sigma}(\tau^{\prime})
\end{eqnarray}
where $S_{atom}$ represents the action of the isolated impurity, and
$\Delta_{\sigma }(\tau-\tau^{\prime})$ is the hybridization function
of the effective AIM.  In this $U=\infty$ case, the double occupied
state of the impurity should be excluded when evaluating $S_{atom}$.
$\Delta_{\sigma }(\tau-\tau^{\prime})$ is not initially known and it
must be determined by the DMFT self-consistency condition given below.
The impurity Green function and the impurity self-energy are given by
the following equations
\begin{eqnarray}
G_{\sigma}(\tau-\tau^{\prime})=-\langle
Tc_{\sigma}(\tau)c^{\dag}_{\sigma}(\tau^{\prime})\rangle_{S_{eff}}\\
\Sigma_{\sigma}(i\omega_{n})=i\omega_{n}+\mu-\Delta_{\sigma}(i\omega_{n})-G_{\sigma}^{-1}(i\omega_{n}).
\end{eqnarray}

The DMFT self-consistency condition requires that the local Green's
function of the lattice coincides with the Green's function of the
auxiliary AIM and identifies the equivalence between the lattice local
self-energy and the self-energy of the corresponding AIM, i.e.,
\begin{eqnarray}\label{scc}
   \sum_{\textbf{k}}\frac{1}{i\omega_{n}+\mu+h\sigma-\epsilon(\textbf{k})-\Sigma_{\sigma}(i\omega_{n})}\nonumber \\
   =\frac{1}{i\omega_{n}+\mu+h\sigma-\Delta_{\sigma}(i\omega_{n})-\Sigma_{\sigma}(i\omega_{n})},
\end{eqnarray}
where the lattice dispersion of our choice is
$\epsilon(\textbf{k})=-2t(\cos k_{x}+\cos k_{y})-4t^{\prime}\cos
k_{x}\cos k_{y}$ and $h$ is the external magnetic field.  For a given
Weiss field $\Delta_{\sigma}(i\omega_{n})$, the effective action
$S_{eff}$ is constructed and the AIM is solved for the new
$G_{\sigma}(i\omega_{n})$ and $\Sigma_{\sigma}(i\omega_{n})$. Using
the self-consistency condition Eq.5, the new Weiss field
$\Delta_{\sigma}(i\omega_{n})$ is computed.
This iterative procedure is repeated until the Green's function is
converged.

To solve the impurity problem of Eq.~\ref{action}, the CTQMC impurity
solver is used.  In this method, the hybridization part of the
effective action is treated as a perturbation around the atomic action
and all diagrams are summed up by stochastic Metropolis
sampling.~\cite{Haule:06} In this $U=\infty$ case, doubly occupied
state of the atom is excluded from atomic eigenstates. CTQMC converges
well in the low Matsubara frequency region, but it is poorly behaved
in the high frequency region. Therefore, one needs the analytic
expression for the self-energy in the high frequency limit and it has
to be interpolated to the low frequency region. The high frequency
expansion for the $U=\infty$ Hubbard model gives
\begin{eqnarray}
    Re[\Sigma_{\sigma}(\infty)]=m_{1\sigma}/m_{0\sigma}^{2}+\mu\\
    Im[\Sigma_{\sigma}(\infty)]=(1-1/m_{0\sigma})\omega
\end{eqnarray}
where
$m_{0\sigma}=\langle\{c_{\sigma},c_{\sigma}^{\dag}\}\rangle=1-n_{-\sigma}$,
$m_{1\sigma}=\langle\{[c_{\sigma},H],c_{\sigma}^{\dag}\}\rangle=-\mu(1-n_{-\sigma})-Tr[\Delta_{-\sigma} G_{-\sigma}]$.
Note the appearance of the kinetic energy $Tr[\Delta_{-\sigma}
  G_{-\sigma}]$ in this expansion which is absent for finite $U$.

Within CTQMC, various spin dependent physical quantities can be
calculated such as occupation numbers
($n_{\uparrow}$,$n_{\downarrow}$) and the local magnetic
susceptibility ($\chi_{loc}$). The $q=0$ magnetic susceptibility of a
lattice can be calculated from $\chi_{loc}$ by evaluating the two
particle vertex functions, which is a numerically demanding task. To
circumvent this difficulty, $\chi_{q=0}$ of a lattice can be
calculated from the ratio of magnetization to the external magnetic
field ($\chi=\frac{dm}{dh}|_{h=0}$).  The external field $h$ alters
the effective action (Eq.~\ref{action}) by adding $h\sigma$ to atomic
energies and the self-consistency condition (Eq.~\ref{scc}) is
enforced to include the spin dependent $h\sigma$ term during DMFT
iterations.  The exclusion of the double occupancy ($U=\infty$)
implies the Hubbard potential energy to vanish and the only relevant
energy is the kinetic energy. The latter is given by
$Tr[\Delta_{\sigma} G_{\sigma}]$, and it is related to the average of
the perturbation order $k$ as follows:
\begin{eqnarray}
   E_{kin,\sigma}=Tr[\Delta_{\sigma} G_{\sigma}]=-T\langle k_{\sigma}\rangle
\end{eqnarray}
where $T$ is temperature. Therefore, it is possible to calculate the
kinetic energy to high accuracy by evaluating $\langle
k_{\sigma}\rangle$.  The free energy, $F$, can also be derived from
the kinetic energy as long as the system is in the Fermi liquid
regime
\begin{eqnarray}\label{F}
   F(T)\cong E_{kin}-\frac{\pi^{2}}{3}Z^{-1}\rho_{0}(\mu)T^{2},
\end{eqnarray}
where $Z$ is the renormalization residue and $\rho_{0}$ is the
non-interacting density of states.

\begin{figure}[!ht]
\centering{
  \includegraphics[width=\linewidth,clip=]{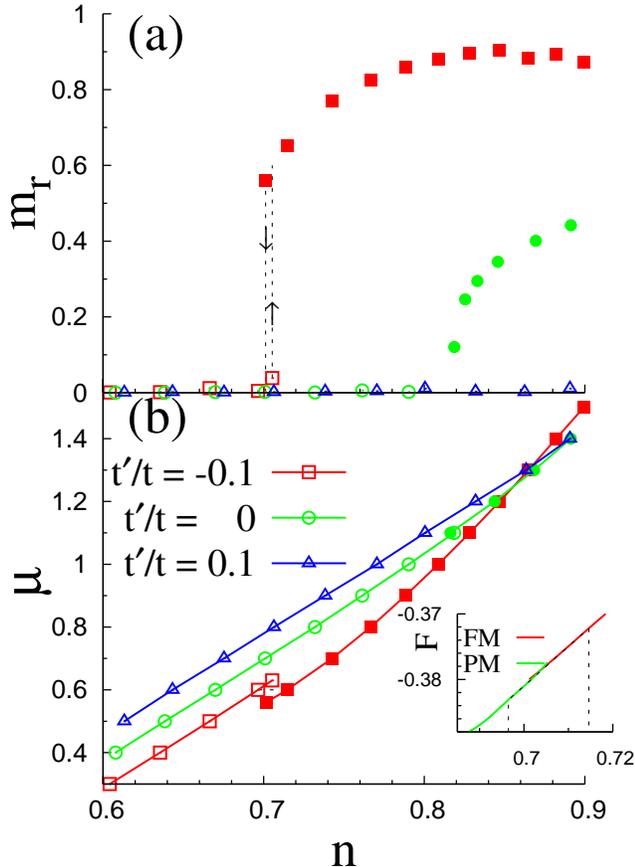}\\
  }
\caption{\label{DMFT} (Color online) (a) The reduced magnetization $m_{r}$=($n_{\uparrow}-n_{\downarrow})/(n_{\uparrow}+n_{\downarrow}$) vs the electron density $n$ at $t^{\prime}/t$=-0.1, 0, and 0.1 (b) the chemical potential $\mu$ vs $n$ at $t^{\prime}/t$=-0.1, 0, and 0.1. Filled points indicate a FM state. Inset : FM free energy and PM free energy vs $n$ at $t^{\prime}/t$=-0.1. The dotted line is constructed using the Maxwell construction.
All calculations were performed at $T$=0.01.}
\end{figure}

\begin{figure}[!ht]
\centering{
  \includegraphics[width=\linewidth,clip=]{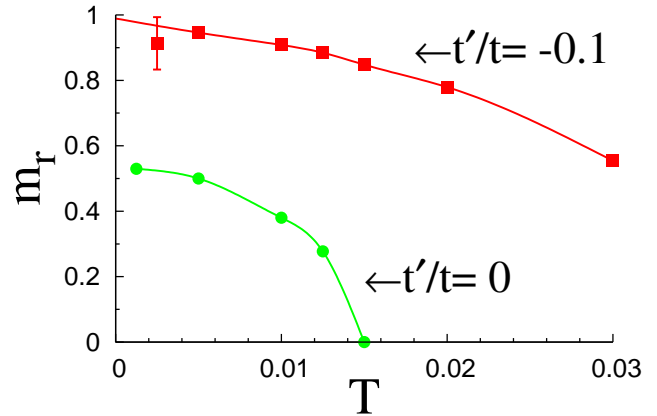}\\
  }
\caption{\label{mvsT} (Color online) $m_{r}$ vs $T$ at fixed $n$ =0.85 with $t^{\prime}/t$=-0.1 and 0.
The fully polarized FM state ($m_{r}=1$) is expected only when $t^{\prime}/t$=-0.1.}
\end{figure}

\begin{figure}[!ht]
\centering{
  \includegraphics[width=\linewidth,clip=]{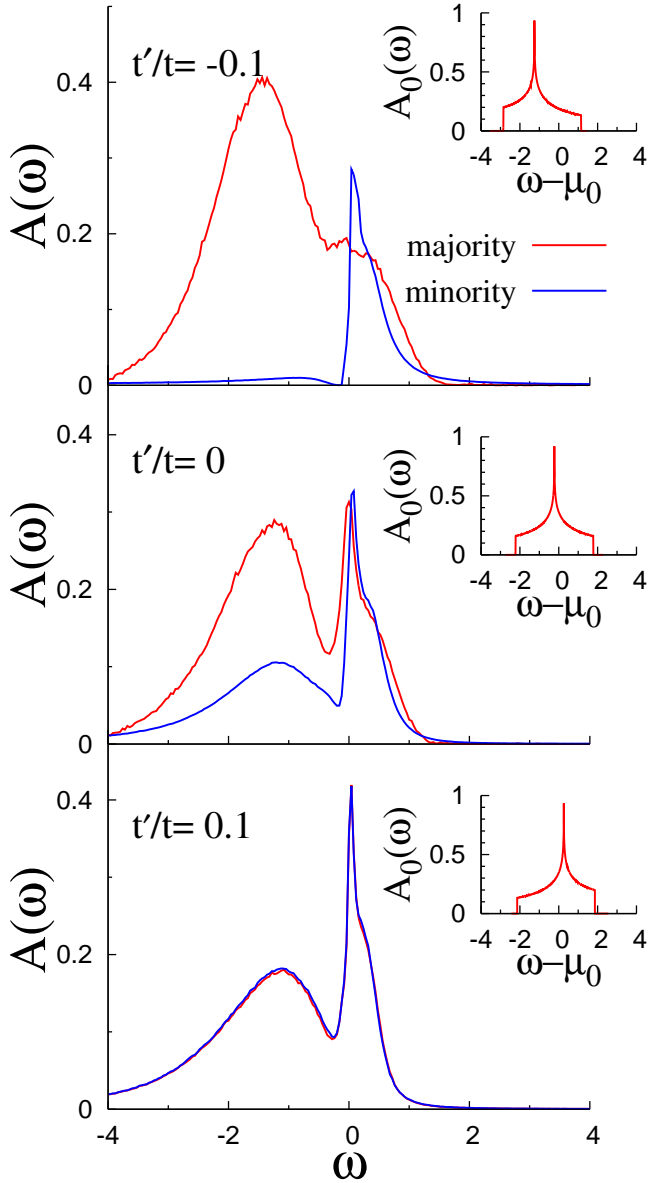}\\
  }
\caption{\label{spectral} (Color online) The spectral functions $A(\omega)$ at $t^{\prime}/t$=-0.1 (top), 0 (middle), and 0.1 (bottom) for fixed $n$ =0.85.
Inset: Non-interacting spectral functions ($A_{0}(\omega)$) of the majority spin at the corresponding $t^{\prime}/t$ values. ($\mu_{0}=\mu-Re\Sigma(0)$)
All calculations were performed at $T$=0.01.}
\end{figure}

\begin{figure}[!ht]
\centering{
  \includegraphics[width=\linewidth,clip=]{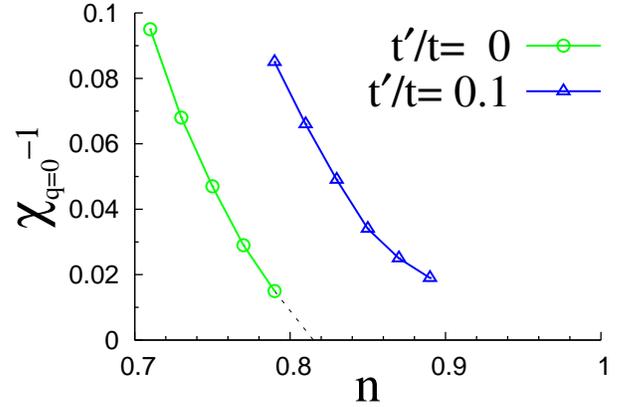}\\
  }
\caption{\label{inv_sus} (Color online) The uniform susceptibility ($\chi_{q=0}^{-1}$) vs $n$ at $t^{\prime}/t=0$ and $0.1$. The dotted line is for the extrapolation to $\chi_{q=0}^{-1}=0$.
($T$=0.01)}
\end{figure}

\begin{figure}[!ht]
\centering{
  \includegraphics[width=\linewidth,clip=]{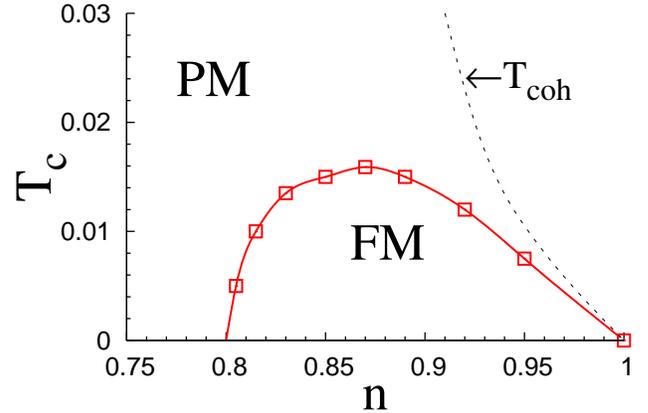}\\
  }
\caption{\label{Tc} (Color online) The critical temperature $T_{c}$ vs $n$ at $t^{\prime}=0$. $n_{c}$ at $T=0$ is obtained from the extrapolation.
The dotted line represents the coherence temperature $T_{coh}$ vs $n$.}
\end{figure}

\begin{figure}[!ht]
\centering{
  \includegraphics[width=\linewidth,clip=]{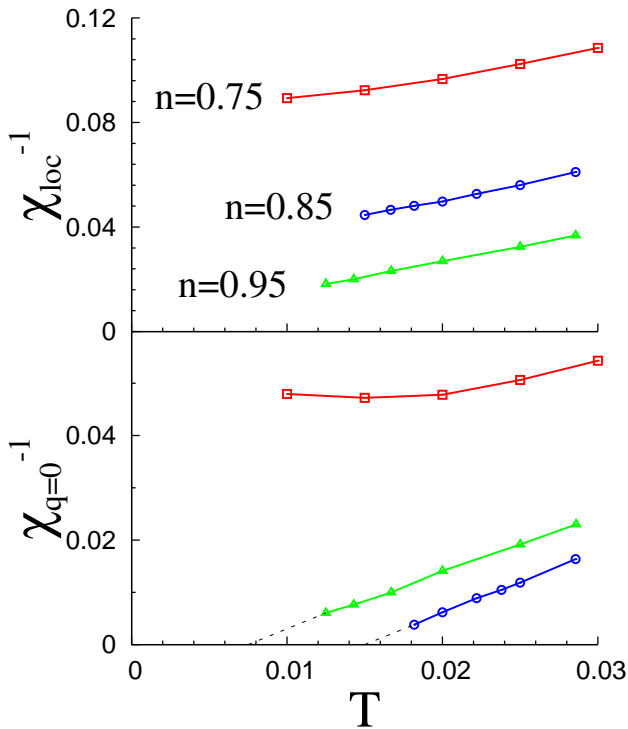}\\
  }
\caption{\label{susvsT} (Color online) The local susceptibility ($\chi_{loc}^{-1}$) vs $T$ and the uniform susceptibility ($\chi_{q=0}^{-1}$) vs $T$ ($t^{\prime}/t=0$)}
\end{figure}

\begin{figure}[!ht]
\centering{
  \includegraphics[width=\linewidth,clip=]{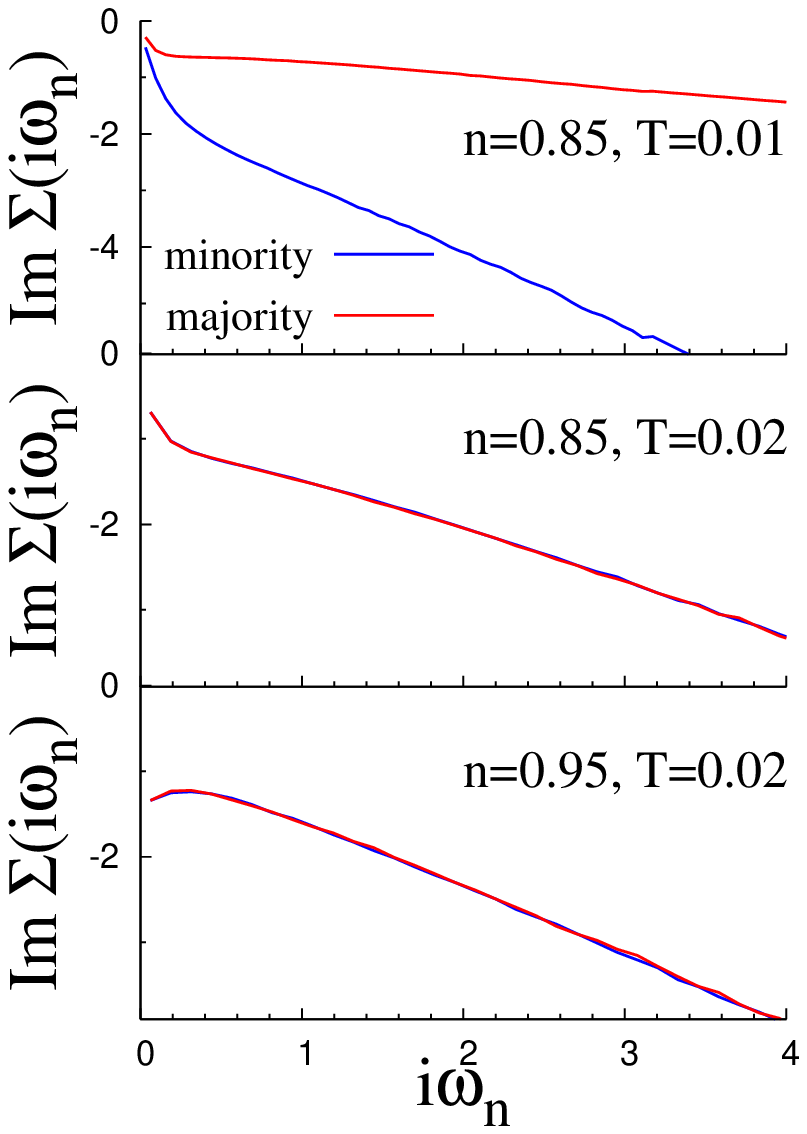}\\
  }
\caption{\label{sigma} (Color online) $Im \Sigma(i\omega_{n}$) vs $i\omega_{n}$ at the coherent FM state (the top panel), the coherent PM state (the middle panel), and the incoherent PM state (the bottom panel). ($t^{\prime}/t=0$)}
\end{figure}

Fig.~\ref{DMFT}.(a) shows the reduced magnetization
$m_{r}$=($n_{\uparrow}-n_{\downarrow})/(n_{\uparrow}+n_{\downarrow}$)
as a function of the electron density $n$ at three distinct
$t^{\prime}/t$ ratios.  The result is notably different for different
values of $t^{\prime}/t$. The spontaneously broken ferromagnetic
(FM) state ($m_{r}\neq0$) is favored for $t^{\prime}/t<0$
while the FM state is unstable for $t^{\prime}/t>0$.
The critical density ($n_{c}$) at which the transition occurs
increases as $t^{\prime}/t$ increases, reducing the region of
stability of the FM state.
Moreover, at $t^{\prime}/t=-0.1$ magnetization $m_{r}$ changes
abruptly at $n_{c}$=0.705 indicating a first order transition, while
at $t^{\prime}/t=0$ magnetization $m_{r}$ increases continuously
indicating a second order phase transition at $n_c$=0.815.

Notice that close to half filling the Curie temperature is low
and at fixed temperature ($T=0.01$) it becomes increasingly difficult
to converge the DMFT equations near the transition temperature due to
the standard critical slowing down.

Near half filling the quasiparticle bandwidth is small due to strong
correlations hence the thermal fluctuations are comparable to the
Curie temperature in this region.  A stable FM state is possible only
if $T$ is sufficiently lower than $T_{coh}$.  In the region above
$0.95$, an incoherent paramagnetic (PM) state becomes stable as $T$
exceeds $T_{coh}$.

Inspecting the chemical potential as a function of density reveals
that the nature of the transition changes with $t^{\prime}/t$ (see
figure \ref{DMFT}.(b)). For $t^{\prime}/t=0$ the transition is
continuous while for $t^{\prime}/t=-0.1$, there is a region of
constant chemical potential which corresponds to a first-order
transition.  The flat chemical potential region ($n=0.696-0.715$)
indicates that two different DMFT solutions (FM, PM) can be converged
depending on the initial conditions and it indicates phase separation
(PS) of the FM and PM state.  This region is determined by Maxwell
construction which connects common tangents between two phases in the
free energy vs $n$ graph. (Fig.~\ref{DMFT}. inset)

The original debate on the Nagaoka problem was focused on the
existence of the fully polarized FM state at finite $\delta$ in the
$T\rightarrow 0$ limit. Therefore, it is necessary to investigate
magnetization $m_{r}$ at very low $T$. In Fig.~\ref{mvsT} we show very
low temperatures ($T=0.001t$) results and it is clear that the
magnetization saturates to a value smaller than unity for
$t^{\prime}/t=0$ while it reaches unity at low temperatures for
$t^{\prime}/t=-0.1$.
The fully polarized Nagaoka state is thus not stable for
$t^{\prime}/t=0$ and moderately small doping ($\delta \sim 0.1$) while
it is realized for $t^{\prime}/t=-0.1$.  As the spins become fully
polarized ($t^{\prime}/t=-0.1, T\rightarrow 0$), numerics requires high
statistics and an error-bar is specified to take into account the
numerical error.

The spectral functions are shown in Fig.~\ref{spectral}. Since CTQMC
delivers response functions on the imaginary frequency axis, one needs
to perform the analytical continuation of the Green function to the
real axis. Here we use the maximum entropy
method~\cite{Jarrell:96}. The spectral functions show noticeable
differences for small change in $t^{\prime}$.
%
%
At $t^{\prime}/t=-0.1$, the majority spin spectral function shows a very
small renormalization due to interactions ($Z\simeq1$) and a large
spectral peak in the occupied part of the spectra. The overall shape
is similar to the non-interacting spectral function
(Fig.~\ref{spectral}.  inset). The minority spin spectral function is
much more correlated and shows a narrow quasiparticle band above the
Fermi level and a tiny lower Hubbard band. In the magnetic state, the
occupied part of the spectra is thus well described by a model of a
weakly correlated FM metal.

At $t^{\prime}/t=0$ and $t^{\prime}/t=0.1$, the spectral functions
consist of both the narrow quasiparticle band and the lower Hubbard
band.  In the $U=\infty$ Hubbard model, the upper Hubbard band
disappears due to the exclusion of double occupancy.

The stability of the FM state at $t^{\prime}/t=-0.1$ can be traced
back to the large spectral peak in the occupied part of the spectra of
the non-interacting DOS shown in the inset of Fig.~\ref{spectral}.  As
explained above, the majority spin of the FM state shows only weak
renormalization due to interactions. This is a consequence of the
Pauli exclusion principle which constrains the motion of a hole in the
polarized background and interactions, being less important in this
case, do not hamper the coherent motion of the hole through the polarized
background. The kinetic energy of this state thus clearly depends on
$t^{\prime}/t$ ratio and is reduced with decreasing $t^\prime/t$.
Contrary to the FM state, the correlations are very strong in the
PM state regardless of the spectral peak in the
non-interacting DOS and $t^{\prime}/t$ ratio. The coherent part of
the spectra does not contribute much to the kinetic energy
as the quasiparticle bandwidth shrinks due to the strong correlations.
The incoherent part of the spectra in the form of the Hubbard bands
arises from localized electrons and consequently
it is almost independent of the specific lattice dispersion.
Therefore, the kinetic energy of the PM state
weakly depends on $t^{\prime}/t$ ratio.
The peak in the occupied part of the spectra of the non-interacting
DOS thus reduces the kinetic energy of the FM
state compared to the PM state thus stabilizing
ferromagnetism.

It is known from other studies \cite{Tasaki:92} that a highly
degenerate flat band in the occupied part of the spectra favors
ferromagnetism at any finite $U$. However, this flat
band ferromagnetism (an extreme limit of the Stoner ferromagnetism)
argument is not applicable to the $t^{\prime}/t=-0.1$ case of the
Nagaoka ferromagnetism (the other extreme limit of the Stoner
ferromagnetism).
In a flat-band model, the ground state of the non-interacting system
is highly degenerate due to the presence of the flat band.
%
%
However, even a small Coulomb repulsion lowers the
energy of the FM state (if the flat band is half-filled)
and stabilizes the FM state.  The role of the Coulomb
interaction is simply to lift the huge degeneracy and "select" the
states with the highest magnetization as unique ground states.
In the infinite $U$ model, the potential energy vanishes because of no
doubly occupancy. However, the kinetic energy depends sensitively on
the smoothness of the spin polarized background, and a disordered
PM state can not gain the kinetic energy by the variation of
$t^\prime/t$ while a FM state can.


The inverse of the uniform magnetic susceptibility ($\chi^{-1}_{q=0}$)
of the PM state vs $n$ at $t^{\prime}/t=0$ and $0.1$ is shown in
Fig.~\ref{inv_sus}.  The extrapolated line at $t^{\prime}=0$ indicates
that $\chi$ diverges near $n=0.815$, confirming the second order
transition at the critical density ($n_{c}=0.815$). At
$t^{\prime}/t=0.1$, one might expect $\chi$ will diverge near
$n=1$. However, as $T_{coh}$ becomes smaller than $T$ near $n=1$, the
incoherent PM state is stabilized. In other words, at
$t^{\prime}/t=0.1$, the crossover from the coherent PM state to the
incoherent PM state occurs instead of the transition to the FM state.

Fig.~\ref{Tc} shows the critical temperature ($T_{c}$) vs $n$ at
$t^{\prime}/t=0$. In the region below $T_{c}$ a partially polarized FM
state is found, and it is determined by observing $n_{\uparrow}\neq
n_{\downarrow}$ in a CTQMC result.  This graph shows that the lower
critical density ($n_{c}$) at $T=0$ is around 0.8.  At half filling
critical temperature should vanish due to the following reason:
The kinetic energy at half filling is zero in both the PM and
the FM state because of the blocking of charge density. The
entropy of the paramagnet is much larger than the entropy of the
ferromagnet due to the large spin degeneracy of the PM
state. In other words, PM state is thermodynamically stable at any
finite temperature at $n=1$.

As the width of the quasiparticle band becomes smaller near $n=1$, the
coherence temperature $T_{coh}$ is also reduced making it hard to
sustain the quasiparticle coherent band.
At $T>T_{coh}$, the PM state is clearly stabilized.  The $T_{coh}$
boundary can be determined from the imaginary part of self energy ($Im
\Sigma(i\omega_{n})$) on the imaginary frequency axis.  In a coherent
region ($T<T_{coh}$), the renormalization residue $Z$ is well defined
($0<Z<1$) by evaluating the negative slope of $Im \Sigma(i\omega_{n})$
at $\omega=0$ ($Z=(1-\frac{d Im\Sigma}{d
\omega}|_{\omega=0})^{-1}$). However, in the incoherent regime
($T>T_{coh}$), the slope of $Im \Sigma(i\omega_{n})$ at $\omega=0$
becomes positive making the concept of $Z$ ill defined
(Fig.~\ref{sigma}). Therefore, we determined $T_{coh}$ as the
temperature where the slope of the low energy self energy vanishes,
and found that it is almost proportional to $\delta^{3/2}$, in
surprising agreement with the findings of a previous study of doped
Mott-insulator~\cite{Kajoeter:96}.

In a two-dimensional Hubbard model, a long-range magnetic order at a
finite $T$ is prohibited by the Mermin-Wagner theorem. A FM
order is possible only at $T=0$. At any finite $T$, Goldstone modes
disorder the system~\cite{Polyakov:75}, and it results in a
correlation length which is finite but exponentially large in
$T^{-1}$.  DMFT does not capture this behavior. Therefore, $T_{c}$ in
the context of the two dimensional model should be interpreted as an
estimate of the temperature where the correlation length gets very
large.  In higher dimensions, we expect a FM state at low $T$ with the
correct dependence on $t^{\prime}/t$.  The Nagaoka ferromagnetism
study using the dispersion of realistic materials deserves further
investigations since the energy balance between a FM state and a PM
state or the character of the transition is very sensitive to the
details of the lattice structure.

In general, $n_{\uparrow}-n_{\downarrow}$ exhibits small fluctuations
near the boundary of $T_{c}$ due to the finite $T$.  The fluctuations
become especially severe through the transition from the FM state to
the incoherent PM state near $n=1$.  Therefore, the boundary points
can be determined more precisely by examining the temperature
dependence of $\chi_{q=0}^{-1}$ (Fig.~\ref{susvsT}). $\chi_{q=0}^{-1}$
near a transition point obeys the Curie-Weiss form
($\chi_{q=0}^{-1}\sim T-T_{c}$). Both coherent ($n=0.85$) and
incoherent ($n=0.95$) regions show linear dependence of
$\chi_{q=0}^{-1}$ on $T$. The $\chi_{q=0}^{-1}$ for $n=0.75$ barely
depends on $T$, exhibiting Pauli paramagnetic
behavior. $\chi_{loc}^{-1}$ is greater than $\chi_{q=0}^{-1}$ and it
increases as $n$ decreases. This is because in DMFT
$\chi_{loc}^{-1}\sim T+T_{coh}$ and $T_{coh}$ increases as $n$
decreases~\cite{Kajoeter:96}.

Fig.~\ref{sigma} shows the behavior of $Im \Sigma(i\omega_{n})$ for
the three different phases in the $T_{c}$ phase diagram of
Fig.~\ref{Tc}. For $n=0.85$ and $T=0.01$, a coherent FM state is
expected from the phase diagram. A coherent Fermi liquid is validated
by investigating the negative slope of $Im \Sigma(i\omega_{n})$ at
$\omega=0$. The slope for spin $\sigma$ at the high frequency part is
given by $-n_{-\sigma}/1-n_{-\sigma}$ (Eq. 7) and the inequality of
the slope indicates $n_{\uparrow}\neq n_{\downarrow}$ confirming the FM
state. The majority spin state has a smaller slope at high frequency
because $n_{-\sigma}$ of the majority spin is smaller than that of the
minority spin. Also, because the slope of the majority spin at
$\omega=0$ is smaller, $Z$ of the majority spin is larger than that of
the minority spin. This means the quasiparticle band of the minority
spin is strongly renormalized by correlations while the majority spin
state tends to be similar to the non-interacting energy
dispersion. For $n=0.85$ and $T=0.02$, a coherent PM state is
established by observing a negative slope at $\omega=0$ and no spin
symmetry breaking. For $n=0.95$ and $T=0.02$, an incoherent PM state
is expected from the positive slope at $\omega=0$ because the concept
of $Z$ is no longer valid and the application of Fermi liquid theory
fails. Lastly, for fixed $T=0.02$, as $n$ increases from 0.85 to 0.95
the slope at high frequency also increases because $n_{-\sigma}$
increases.

\section{Nagaoka Ferromagnetism from a 4-site plaquette}
\begin{figure}[h]
\centering
  \includegraphics[width=\linewidth,clip= ]{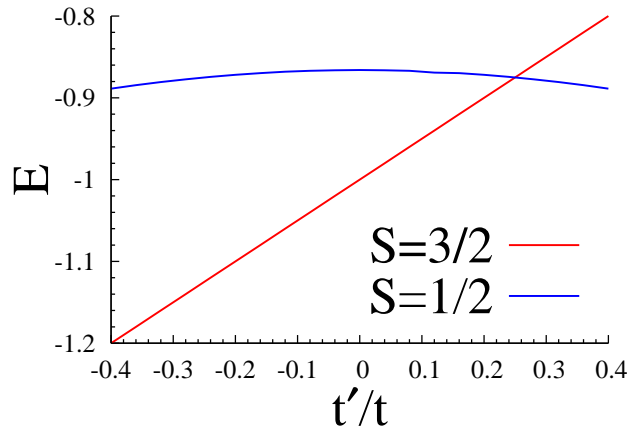}\\
\caption{\label{evstp} (Color online) The lowest energies of a $S$=1/2 state and a
  $S$=3/2 state in a $U=\infty$ 4-site toy model varying
  $t^{\prime}/t$. E is the energy in units of $t=1/2$.  }
\end{figure}

In order to provide a simple interpretation of why decreasing
$t^{\prime}$ stabilizes the Nagaoka state, we examine the simplest
possible model which retains the physics of the Nagaoka problem.  We
consider a 4-site plaquette with three electrons (one hole).  The
ground state of this model may be characterized by the quantum number
corresponding to the total spin angular momentum
(ie. $S=\frac{3}{2},\frac{1}{2}$) and the $z$-direction of the spin
angular momentum ($S_{z}=\pm\frac{3}{2},S_{z}=\pm\frac{1}{2}$).  The
whole Hamiltonian matrix is a $32\times32$ matrix excluding
double-occupied sites and it is block-diagonalized to 6 distinct spin
sectors by performing the unitary transform to the proper $S$, $S_{z}$
basis.  The ground state energy at each spin sector is determined by
the exact diagonalization of Hamiltonian matrix.

The lowest energy in a $S=\frac{3}{2}$ sector is given by
$-2t+t^{\prime}$ and in the $S=\frac{1}{2}$ sector is given by
$-\sqrt{3t^{2}+t^{\prime2}}$.  The energy dependence of a
$S=\frac{3}{2}$ state is noticeably different from that of a
$S=\frac{1}{2}$. In a $S=\frac{3}{2}$ case, doubly occupied states are
excluded by the Pauli principle regardless of $U$. Therefore, the
$U=\infty$ Hamiltonian is equivalent to the $U=0$ Hamiltonian where
the addition of the positive n.n.n hopping $t^{\prime}$ contributes
linearly to the increase of the kinetic energy.  However, doubly
occupied states in a $S=\frac{1}{2}$ sector are excluded only for
$U\rightarrow\infty$. Therefore, unlike the $S=\frac{3}{2}$ case, the
energy dependence on $t^{\prime}$ is greatly reduced as the Hilbert
space shrinks due to the infinite $U$.

A $S=\frac{3}{2}$ ground state is indicative of the Nagaoka
ferromagnetic state while a $S=\frac{1}{2}$ ground state is indicative
of a paramagnetic state.  The $S=\frac{3}{2}$ state is the ground
state for $t^{\prime}/t<0.24$ and the energy difference increases
approximately linearly thereafter indicating that the Nagaoka state is
stabilized as $t^{\prime}/t$ is decreased. This is in qualitative
agreement with the DMFT results presented in the previous section. The
energy of the $S=\frac{1}{2}$ state weakly depends on $t^{\prime}$
while the $S=\frac{3}{2}$ energy decreases as $t^{\prime}/t$
decreases. This also explains that the stability of Nagaoka
ferromagnetism originates from the minimization of the kinetic energy.

\section{A mean-field slave boson approach}

In this section, Nagaoka ferromagnetism in a $U=\infty$ Hubbard model
is studied using a mean-field slave boson approach. In a slave
boson method, a fermion operator is accompanied by bosonic operators
(ie. slave bosons) which keep track of the local occupation
number. The three slave boson operators are
$\hat{e},\hat{p_{\uparrow}},\hat{p_{\downarrow}}$ and they act on
unoccupied sites, spin-up sites, and spin-down sites, respectively. In
this $U=\infty$ case, the doubly occupied sites are excluded.
Constraints regarding the conservation of the occupation number are
imposed with Lagrange multipliers ($\lambda, \lambda_{\uparrow},
\lambda_{\downarrow}$). The slave boson Hamiltonian is given by
\begin{eqnarray}
   \hat{H}=-\sum_{ij\sigma}t_{ij}\hat{c}_{i\sigma}^{\dagger}\hat{z}_{i\sigma}\hat{z}_{j\sigma}^{\dagger}\hat{c}_{j\sigma}-
   \sum_{i\sigma} & \lambda_{i\sigma}(\hat{p}_{i\sigma}^{\dagger}\hat{p}_{i\sigma}-
   \hat{c}_{i\sigma}^{\dagger}\hat{c}_{i\sigma})+ \nonumber \\
   \sum_{i\sigma}\lambda_{i}(\hat{p}_{i\sigma}^{\dagger}\hat{p}_{i\sigma}+\hat{e}_{i}^{\dagger}\hat{e}_{i}-1)
\end{eqnarray}
where
$\hat{z}_{i\sigma}=\frac{1}{\sqrt{1-\hat{p}_{i\sigma}^{\dagger}\hat{p}_{i\sigma}}}\hat{e}_{i}^{\dagger}\hat{p}_{i\sigma}\frac{1}{\sqrt{1-\hat{e}_{i}^{\dagger}\hat{e}_{i}-\hat{p}_{i-\sigma}^{\dagger}\hat{p}_{i-\sigma}}}$.
$t_{ij}$=$t$ if i,j are n.n, and $t_{ij}$=$t^{\prime}$ if i,j are
n.n.n.  The non-interacting $\epsilon(\textbf{k})$ is taken to be
$-2t(\cos k_{x}+\cos k_{y})-4t^{\prime}\cos k_{x}\cos k_{y}$ as in the
previous section.  The original Fock space has been enlarged including
the slave boson fields. The partition function can be calculated from
the Feynman functional path integral over the original fermi fields,
slave boson fields, and Lagrange multipliers. The integral over the
fermi fields is straightforward because the Hamiltonian is quadratic
in the fermi fields.  The integral over the slave boson fields and
Lagrange multipliers should be performed using the saddle-point
approximation, where the integral over the slave boson fields and
Lagrange multipliers is approximated by putting their space and time
independent mean-field values which minimize the Hamiltonian.  The
physical meaning of slave boson mean-field value is clear.  The
expectation value $\langle \hat{e}^{\dagger}\hat{e}\rangle$
corresponds to the fraction of unoccupied sites, i.e.  the hole
density $\delta (1-n)$.  Similarly, $\langle
\hat{p}_{\uparrow}^{\dagger}\hat{p}_{\uparrow}\rangle$ equals to the
spin up occupation number ($n_{\uparrow}$), and $\langle
\hat{p}_{\downarrow}^{\dagger}\hat{p}_{\downarrow}\rangle$ corresponds
to the spin down occupation number ($n_{\downarrow}$).

The free energy can be derived from the partition function
($F=-k_{B}T\ln Z$) and it is necessary to compare the free energies
between ferromagnetic state and paramagnetic state to investigate the
transition. The free energy is a function of magnetization $m =
n_{\uparrow}-n_{\downarrow}$, $\delta$, and $T$.  At $T=0$, the free
energy becomes the ground state energy. The energies of the fully
polarized ferromagnetic (FPFM) state ($m=n_{\uparrow}$) and the
paramagnetic (PM) state ($m=0$) are given by.
\begin{eqnarray}
   E_{FPFM}(\delta)=\frac{1}{N_{s}}\sum_{\textbf{k}}\epsilon(\textbf{k})\Theta(\mu-\epsilon(\textbf{k}))\\
   E_{PM}(\delta)=\frac{1}{N_{s}}\sum_{\textbf{k},\sigma}Z\epsilon(\textbf{k})\Theta(\mu^{*}-Z\epsilon(\textbf{k}))\label{slavebosonE}
\end{eqnarray}
where $N_{s}$ is the number of total sites, $Z$ is the renormalization
residue given by $2\delta/(1+\delta)$, $\mu$ is the chemical potential
in a fully polarized ferromagnetic state satisfying
$(1/N_{s})\sum_{\textbf{k}}\Theta(\mu-\epsilon(\textbf{k}))=n_{\uparrow}=1-\delta$,
and $\mu^{*}=(\mu-\lambda_{\sigma})$ is the effective chemical
potential in a paramagnetic state satisfying
$(1/N_{s})\sum_{\textbf{k}}\Theta(\mu^{*}-Z\epsilon(\textbf{k}))=n_{\uparrow}=n_{\downarrow}=(1-\delta)/2$.
The DOS of the FPFM state is the same as the non-interacting DOS
($\rho_{0}(\epsilon)$) while the DOS of the PM state is renormalized
by a factor $Z$ to $1/Z\cdot\rho_{0}(\epsilon/Z)$. Unlike the DMFT
method, the slave boson approach considers only the renormalized
quasiparticle DOS ignoring the incoherent contribution.  $E_{PM}$ is
given by $Z\cdot E_{0}$ where $E_{0}$ is the non-interacting
energy. In other words, as $\delta$ reduces to 0, the energy for a
paramagnetic state is strongly renormalized by a factor
$2\delta/(1+\delta)$ to avoid the doubly occupied states. That makes
the FPFM state more stable at small $\delta$.

\begin{figure}[!ht]
\centering
  \includegraphics[width=\linewidth,clip= ]{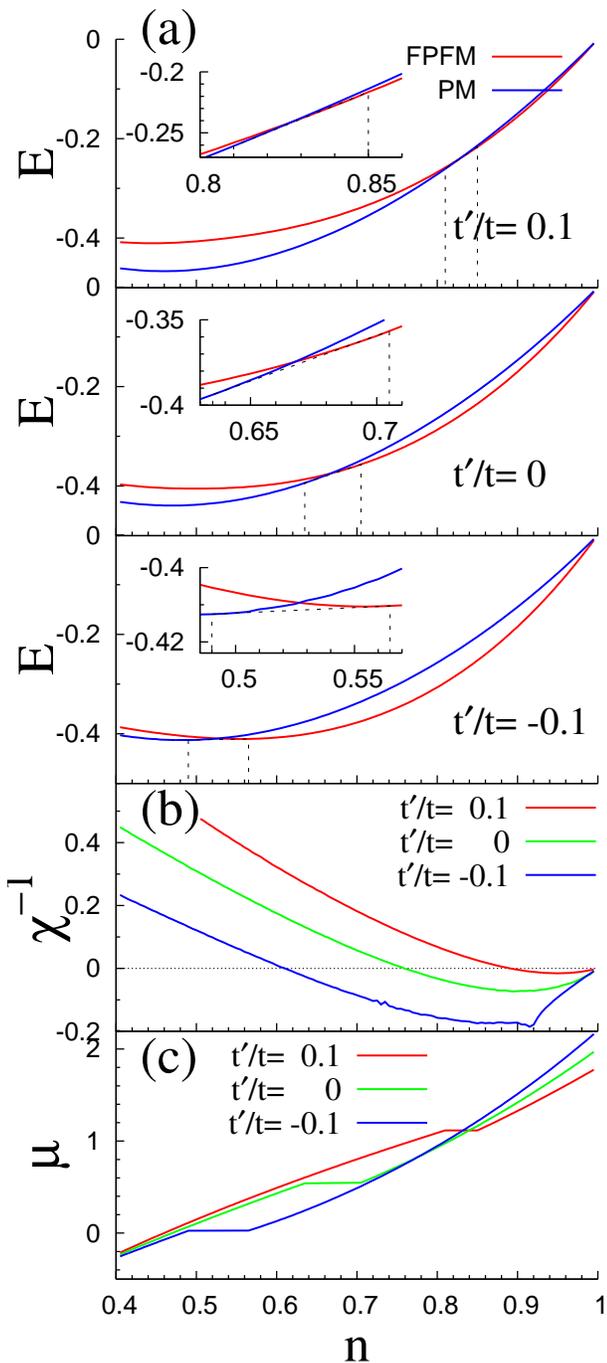}
\caption{\label{slaveboson} (Color online) (a) Fully polarized ferromagnetic (FPFM)
energy and paramagnetic (PM) energy vs $n$ varying $t^{\prime}/t$ (0.1 (top), 0 (middle), and -0.1 (bottom))
Inset : Maxwell construction to determine the PS region.
(b) The inverse of the uniform magnetic susceptibility ($\chi^{-1}$) at $m$=0 vs $n$ varying
$t^{\prime}/t$ (0.1, 0, and -0.1). (c) The chemical potential ($\mu$) vs $n$ at
$t^{\prime}/t$ = 0.1, 0, and -0.1.}
\end{figure}

In Fig.~\ref{slaveboson}.(a), FPFM
energy and PM energy vs $n$ are shown for $t^{\prime}/t=0.1$, 0 and
$-0.1$.  For all values of $t^{\prime}/t$, the FPFM energy is stable
at large $n$ while the PM energy is stable at small $n$. The
intermediate phase separated region is constructed by the Maxwell
construction and is indicative of a fist order transition.  At large
$n$, as in the plaquette case, the energy curve for the paramagnet
state depends weakly on $t^{\prime}$ while the FPFM energy is reduced
with decreasing $t^{\prime}/t$.
This results is in qualitative agreement with the previous DMFT
results.  As $t^{\prime}/t$ decreases, the FPFM state becomes more
stable and the critical density, $n_{c}$ decreases.  Just as in the
DMFT, the large spectral weight of the non-interacting DOS at a low
energy makes FPFM state energetically favorable at
$t^{\prime}/t=-0.1$.
When $t^{\prime}$ is 0, the energy difference between FPFM and PM
vanishes at $n_{c}$=2/3, in agreement with the previous slave boson
calculations~\cite{Moller:93,Boies:95}.

We also calculate the inverse of uniform magnetic susceptibility ($\chi^{-1}$)
to study the instability of the PM state. The analytic expression is
\begin{eqnarray}
   \chi^{-1}|_{m=0}=\frac{1}{2\rho(\mu^{*})}+
   \frac{2\mu^{*}}{1+\delta}+ \nonumber \\
   \frac{1}{N_{s}}\sum_{\textbf{k}}\frac{4}{(1+\delta)^{2}}Z\epsilon(\textbf{k})\Theta(\mu^{*}-Z\epsilon(\textbf{k}))
\end{eqnarray}
where $\rho(\mu^{*})$ is the renormalized DOS given by $1/Z\cdot\rho_{0}(\mu^{*}/Z)$.

The trends in $\chi^{-1}$ are consistent with the results shown in
Fig.~\ref{slaveboson}.(a).
As $t^{\prime}/t$ decreases, spin susceptibility diverges at smaller
density (see Fig.~\ref{slaveboson} (b)).
However, the divergence of the spin susceptibility does not coincide
with the thermodynamic phase transition identified by the total energy
differences. The phase transition is thus always first order within
the slave boson approach.
%
%

Fig.~\ref{slaveboson} (c) shows that a flat chemical potential region
exists at any $t^{\prime}/t$s in a $\mu$ vs $n$ graph.  This is a
generic feature of a first order transition and this region represents
the coexistence of the FPFM and PM phase.  This coexistence region is
larger for negative $t^{\prime}/t$ favoring transition to the FPFM
phase.

\section{Comparison of the slave boson result and the DMFT+CTQMC result}
\begin{table}
    \begin{tabular}{|c|c|c|c|c|}
        \hline
        & & $t^{\prime}/t$=-0.1 & $t^{\prime}/t$=0 & $t^{\prime}/t$=0.1 \\
        \hline
        DMFT & $n_{c}$  & 0.705 & 0.815 & N/A \\
        \cline{2-5}
        ($T=0.01$) & order & First & Second & N/A \\
        \hline
        Slave boson & $n_{c}$  & 0.53 & 0.67 & 0.83 \\
        \cline{2-5}
        ($T=0$) & order & First & First & First \\
        \hline
    \end{tabular}
  \caption{$n_{c}$ and the order of the ferromagnetism transition
  in a $U=\infty$ Hubbard model from both the DMFT+CTQMC
  approach and the slave boson approach with $t^{\prime}/t$= -0.1, 0, and 0.1.
  N/A means no transition to FM state occurs.}
  \end{table}

\begin{figure}[!ht]
\centering
  \includegraphics[width=\linewidth,clip= ]{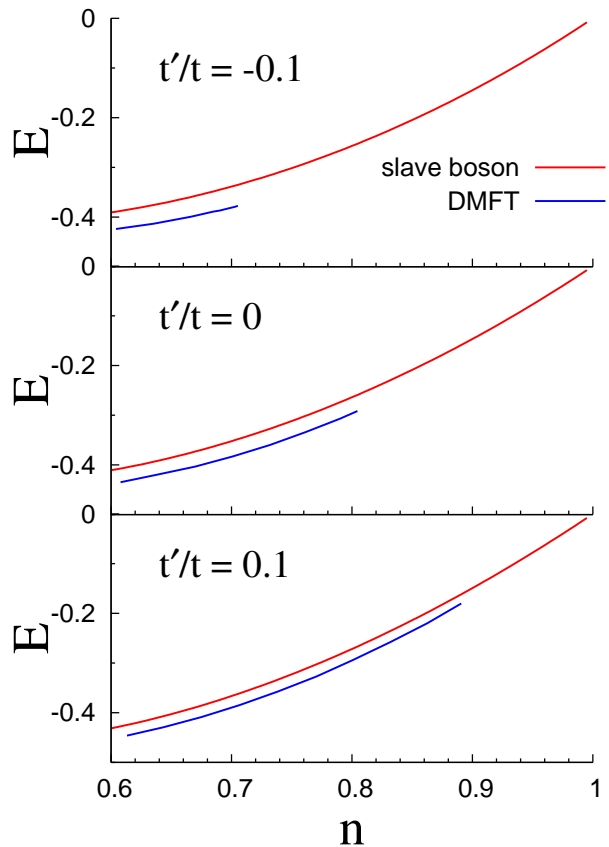}\\
\caption{\label{PMenergy} (Color online) Paramagnetic energy from both the DMFT+CTQMC ($T=0.01$) and the slave boson ($T=0$) approach vs $n$
at $t^{\prime}/t$=-0.1 (the top panel), 0 (the middle panel), and 0.1 (the bottom panel).}
\end{figure}

\begin{figure}[!ht]
\centering
  \includegraphics[width=\linewidth,clip= ]{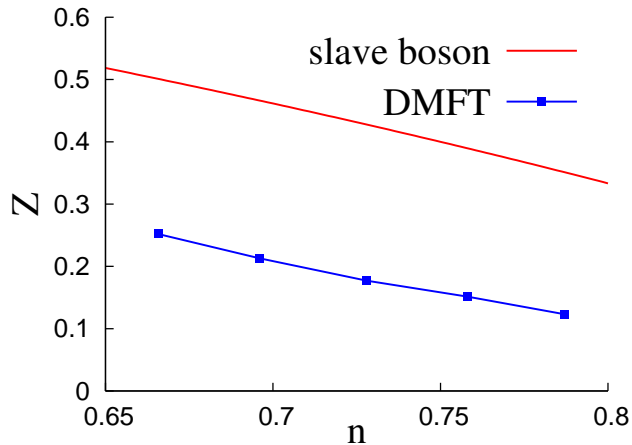}\\
\caption{\label{Zvsn} (Color online) The renormalization residue ($Z$) of the slave boson method
and the DMFT+CTQMC method ($t^{\prime}=0$).
}
\end{figure}

The slave boson method overestimates the region of the stable
FM state as compared to DMFT and it favors a first order
transition (see Table 1).
This is because the slave boson approach overestimates the
paramagnetic kinetic energy as compared to the DMFT approach
(Fig.~\ref{PMenergy}).  The quasiparticle residue $Z$ of the DMFT
approach is evaluated by $(1-\frac{d Im\Sigma}{d
\omega}|_{\omega=0})^{-1}$ on the imaginary frequency axis while $Z$
of the slave boson approach is given by $2\delta/(1+\delta)$.
Fig.~\ref{Zvsn} shows that $Z$ of the slave boson study is
overestimated as compared to the DMFT+CTQMC case.  The slave boson
technique used in this paper is based on the mean-field saddle-point
approximation and it does not treat the strong correlation effect
properly.  Even though DMFT ignores the spatial correlation effect,
the temporal correlations are treated exactly by CTQMC.  Moreover, the
mean-field slave boson approach evaluates the total energy as the sum
of coherent quasiparticle energies (Eq.~\ref{slavebosonE}) while the
total energy of DMFT+CTQMC includes contributions from both the
incoherent and coherent effects.  The over-estimated $Z$ in the slave
boson case underestimates the kinetic energy while the ignorance of
contribution from the incoherent part overestimates the energy.  As a
result, the two errors of the slave boson approach cancel each other
giving a slightly overestimated energy as compared to the DMFT+CTQMC
result.

Additionally, the $\chi^{-1}$ graph in the slave boson method almost
coincides with the DMFT+CTQMC result comparing Fig.~\ref{inv_sus} and
Fig.~\ref{slaveboson} (b). It is not certain how the renormalization
residue $Z$ affects $\chi^{-1}$ in the DMFT+CTQMC case, and the
contribution from the incoherent part is also unclear.  Therefore,
further study will be required to fully understand the positive
agreement of $\chi$ in the two methods.

\section{Conclusion}

To summarize, we investigated Nagaoka ferromagnetism in the $U=\infty$
Hubbard model including n.n hopping $t$ and n.n.n hopping
$t^{\prime}$.  This model was solved using DMFT with CTQMC, and the
mean-field slave boson approach.  Even a small value of $t^{\prime}/t$
yields a significant impact on the stability of Nagaoka ferromagnetism.
The DMFT results show that the FM state is more stable for
negative $t^{\prime}/t$, and this is supported by the slave boson
method (see Table 1) and can also be understood from diagonalization
of the 4-site plaquette.  The energy of the minimum spin state ($S=1/2$)
depends weakly on $t^{\prime}/t$, while the energy of the maximum spin
state ($S=3/2$) depends linearly on $t^{\prime}/t$.  Therefore, the
maximum spin state becomes more stable for negative $t^{\prime}/t$.

In both slave boson and DMFT methods, the paramagnetic energy does not
vary much with $t^{\prime}/t$ due to the strong renormalization of the
quasiparticle band (see Fig.~\ref{PMenergy}).  However, the fully
polarized ferromagnetic energy depends on $t^{\prime}/t$ in a similar
fashion as the non-interacting kinetic energy since the correlations are
weaker in the broken symmetry state.
The negative $t^{\prime}/t$ gives a high spectral peak in the occupied
part of the spectra of the non-interacting system.  As a result, the
energy of the FM state is lower and the ferromagnetism is stabilized
in this case.

Within DMFT, the nature of the transition also varies with
$t^{\prime}/t$.  A first order transition accompanied by the PS of the
FM and PM state occurs at $t^{\prime}/t=-0.1$ while a second order
transition occurs at $t^{\prime}/t=0$.  In the slave boson approach,
the transition is always first order regardless of $t^{\prime}/t$.
This is because the slave boson method overestimates the PM energy.
The DMFT result shows that when $n \rightarrow 1$, the FM state
becomes unstable as $T$ exceeds $T_{coh}$.  In other words,
ferromagnetic state is only stable within the coherent Fermi liquid
regime.

The $U=\infty$ one band Hubbard model is a toy model and does not
describe any specific material. However it is physically realizable in
an optical lattice, due to the recent developments in controlling cold
atoms in optical traps \cite{Jaksch:05,Georges:07}. These systems are
highly tunable, and the hopping parameter $t$ and the on-site
interaction $U$ can be adjusted by varying the ratio of the potential
depth of the optical lattice to the recoil energy ($V_{0}/E_{R}$) or
the ratio of interatomic scattering length to the lattice spacing
($a_{s}/d$).  In order to realize the one-band Hubbard model with a
large $U$ ($U/t\geq 100$), $V_{0}/E_{R} \approx 30$ and $a_{s}/d\leq
0.01$ should be the range of parameters in the optical lattice (See
Fig. 4 of Ref. 29).  The tuning of the next-nearest neighbor hopping
$t^{\prime}$ can be achieved by engineering optical lattices with a
non-separable laser potential over each coordinate axis.

It will be very interesting to test these DMFT results experimentally.
Usually, the atomic trap potential is applied to confine atoms in the
optical lattice, and the potential varies smoothly having the minimum
at the center of the trap.  The phase separation between the FM and
the PM phase at $t^{\prime}/t=-0.1$ (taking place between the
densities $n=0.696-0.715$) can be observed in the optical lattice as
three spatially separated distinct regions.  The atom-rich FM region
will tend to move to the center of the optical lattice to be
energetically stabilized while the hole-rich PM region will reside on
the edge of the optical lattice.  Since the total spin is a conserved
quantity, the FM region will be located at the center of the
trap and will consist of two domains containing the up or down
species.  Raising the temperature will destroy the ferromagnetic
magnetic state and consequently the spatial patterns within the trap.

\section*{ACKNOWLEDGMENT}
This research was supported by NSF grant No. DMR 0528969.  Hyowon Park
acknowledges support from the Korea Science and Engineering Foundation
Grant funded by the Korea government (MOST) (No. KRF-2005-215-C00050)

\bibliography{main}

\begin{thebibliography}{29}
\expandafter\ifx\csname natexlab\endcsname\relax\def\natexlab#1{#1}\fi
\expandafter\ifx\csname bibnamefont\endcsname\relax
  \def\bibnamefont#1{#1}\fi
\expandafter\ifx\csname bibfnamefont\endcsname\relax
  \def\bibfnamefont#1{#1}\fi
\expandafter\ifx\csname citenamefont\endcsname\relax
  \def\citenamefont#1{#1}\fi
\expandafter\ifx\csname url\endcsname\relax
  \def\url#1{\texttt{#1}}\fi
\expandafter\ifx\csname urlprefix\endcsname\relax\def\urlprefix{URL }\fi
\providecommand{\bibinfo}[2]{#2}
\providecommand{\eprint}[2][]{\url{#2}}

\bibitem[{\citenamefont{Nagaoka}(1966)}]{Nagaoka:66}
\bibinfo{author}{\bibfnamefont{Y.}~\bibnamefont{Nagaoka}},
  \bibinfo{journal}{Phys. Rev.} \textbf{\bibinfo{volume}{147}},
  \bibinfo{pages}{392} (\bibinfo{year}{1966}).

\bibitem[{\citenamefont{Fazekas}(1999)}]{Patrick}
For recent reviews see : P. Fazekas, {\textit{Lecture Notes on Electron Correlation and Magnetism}, Series in Modern Condensed Matter Physics - Vol. 5} 
(World Scientific, 1999)

\bibitem[{\citenamefont{Shastry et~al.}(1990)\citenamefont{Shastry,
  Krishnamurthy, and Anderson}}]{Shastry:90}
\bibinfo{author}{\bibfnamefont{B.S.}~\bibnamefont{Shastry}},
  \bibinfo{author}{\bibfnamefont{H.R.}~\bibnamefont{Krishnamurthy}},
  \bibnamefont{and} \bibinfo{author}{\bibfnamefont{P.W.}~\bibnamefont{Anderson}},
  \bibinfo{journal}{Phys. Rev. B} \textbf{\bibinfo{volume}{41}},
  \bibinfo{pages}{2375 } (\bibinfo{year}{1990}).

\bibitem[{\citenamefont{Basile and Elser}(1990)}]{Basile:90}
\bibinfo{author}{\bibfnamefont{A.G.}~\bibnamefont{Basile}} \bibnamefont{and}
  \bibinfo{author}{\bibfnamefont{V.}~\bibnamefont{Elser}},
  \bibinfo{journal}{Phys. Rev. B} \textbf{\bibinfo{volume}{41}},
  \bibinfo{pages}{4842 } (\bibinfo{year}{1990}).

\bibitem[{\citenamefont{von~der Linden and Edwards}(1991)}]{Linden:91}
\bibinfo{author}{\bibfnamefont{W.}~\bibnamefont{von~der Linden}}
  \bibnamefont{and} \bibinfo{author}{\bibfnamefont{D.}~\bibnamefont{Edwards}},
  \bibinfo{journal}{J. Phys. Condens. Matter.} \textbf{\bibinfo{volume}{3}},
  \bibinfo{pages}{4917 } (\bibinfo{year}{1991}).

\bibitem[{\citenamefont{Hanisch and Muller-Hartmann}(1993)}]{Hanisch:93}
\bibinfo{author}{\bibfnamefont{T.}~\bibnamefont{Hanisch}} \bibnamefont{and}
  \bibinfo{author}{\bibfnamefont{E.}~\bibnamefont{Muller-Hartmann}},
  \bibinfo{journal}{Ann. Phys} \textbf{\bibinfo{volume}{2}},
  \bibinfo{pages}{381 } (\bibinfo{year}{1993}).

\bibitem[{\citenamefont{Wurth et~al.}(1996)\citenamefont{Wurth, Uhrig, and
  Mueller-Hartmann}}]{Wurth:96}
\bibinfo{author}{\bibfnamefont{P.}~\bibnamefont{Wurth}},
  \bibinfo{author}{\bibfnamefont{G.}~\bibnamefont{Uhrig}}, \bibnamefont{and}
  \bibinfo{author}{\bibfnamefont{E.}~\bibnamefont{Mueller-Hartmann}},
  \bibinfo{journal}{Ann. Phys.} \textbf{\bibinfo{volume}{5}},
  \bibinfo{pages}{148 } (\bibinfo{year}{1996}).

\bibitem[{\citenamefont{Moller et~al.}(1993)\citenamefont{Moller, Doll, and
  Fresard}}]{Moller:93}
\bibinfo{author}{\bibfnamefont{B.}~\bibnamefont{Moller}},
  \bibinfo{author}{\bibfnamefont{K.}~\bibnamefont{Doll}}, \bibnamefont{and}
  \bibinfo{author}{\bibfnamefont{R.}~\bibnamefont{Fresard}},
  \bibinfo{journal}{J. Phys., Condens. Matter.} \textbf{\bibinfo{volume}{5}},
  \bibinfo{pages}{4847 } (\bibinfo{year}{1993}).

\bibitem[{\citenamefont{Boies et~al.}(1995)\citenamefont{Boies, Jackson, and
  Tremblay}}]{Boies:95}
\bibinfo{author}{\bibfnamefont{D.}~\bibnamefont{Boies}},
  \bibinfo{author}{\bibfnamefont{F.}~\bibnamefont{Jackson}}, \bibnamefont{and}
  \bibinfo{author}{\bibfnamefont{A.-M.} \bibnamefont{Tremblay}},
  \bibinfo{journal}{Int. J. Mod. Phys. B} \textbf{\bibinfo{volume}{9}},
  \bibinfo{pages}{1001 } (\bibinfo{year}{1995}).

\bibitem[{\citenamefont{Zhang et~al.}(1991)\citenamefont{Zhang, Abrahams, and
  Kotliar}}]{Zhang:91}
\bibinfo{author}{\bibfnamefont{X.Y.}~\bibnamefont{Zhang}},
  \bibinfo{author}{\bibfnamefont{E.}~\bibnamefont{Abrahams}}, \bibnamefont{and}
  \bibinfo{author}{\bibfnamefont{G.}~\bibnamefont{Kotliar}},
  \bibinfo{journal}{Phys. Rev. Lett.} \textbf{\bibinfo{volume}{66}},
  \bibinfo{pages}{1236 } (\bibinfo{year}{1991}).

\bibitem[{\citenamefont{Becca and Sorella}(2001)}]{Becca:06}
\bibinfo{author}{\bibfnamefont{F.}~\bibnamefont{Becca}} \bibnamefont{and}
  \bibinfo{author}{\bibfnamefont{S.}~\bibnamefont{Sorella}},
  \bibinfo{journal}{Phys. Rev. Lett.} \textbf{\bibinfo{volume}{86}},
  \bibinfo{pages}{3396 } (\bibinfo{year}{2001}).

\bibitem[{\citenamefont{Oles and Prelovsek}(1991)}]{Oles:91}
\bibinfo{author}{\bibfnamefont{A.M.}~\bibnamefont{Oles}} \bibnamefont{and}
  \bibinfo{author}{\bibfnamefont{P.}~\bibnamefont{Prelovsek}},
  \bibinfo{journal}{Phys. Rev. B} \textbf{\bibinfo{volume}{43}},
  \bibinfo{pages}{13348 } (\bibinfo{year}{1991}).

\bibitem[{\citenamefont{Tasaki}(1992)}]{Tasaki:92}
\bibinfo{author}{\bibfnamefont{H.}~\bibnamefont{Tasaki}},
  \bibinfo{journal}{Phys. Rev. Lett.} \textbf{\bibinfo{volume}{69}},
  \bibinfo{pages}{1608 } (\bibinfo{year}{1992}).

\bibitem[{\citenamefont{Gagliano et~al.}(1990)\citenamefont{Gagliano, Bacci, and
  Dagotto}}]{Gagliano:90}
\bibinfo{author}{\bibfnamefont{E.}~\bibnamefont{Gagliano}},
  \bibinfo{author}{\bibfnamefont{S.}~\bibnamefont{Bacci}}, \bibnamefont{and}
  \bibinfo{author}{\bibfnamefont{E.}~\bibnamefont{Dagotto}},
  \bibinfo{journal}{Phys. Rev. B} \textbf{\bibinfo{volume}{42}},
  \bibinfo{pages}{6222 } (\bibinfo{year}{1990}).

\bibitem[{\citenamefont{Hanisch et~al.}(1997)\citenamefont{Hanisch, Uhrig, and
  Muller-Hartmann}}]{Hanisch:97}
\bibinfo{author}{\bibfnamefont{T.}~\bibnamefont{Hanisch}},
  \bibinfo{author}{\bibfnamefont{G.S.}~\bibnamefont{Uhrig}}, \bibnamefont{and}
  \bibinfo{author}{\bibfnamefont{E.}~\bibnamefont{Muller-Hartmann}},
  \bibinfo{journal}{Phys. Rev. B} \textbf{\bibinfo{volume}{56}},
  \bibinfo{pages}{13960 } (\bibinfo{year}{1997}).

\bibitem[{\citenamefont{Wahle et~al.}(1998)\citenamefont{Wahle, Blumer,
  Schlipf, Held, and Vollhardt}}]{Wahle:98}
\bibinfo{author}{\bibfnamefont{J.}~\bibnamefont{Wahle}},
  \bibinfo{author}{\bibfnamefont{N.}~\bibnamefont{Blumer}},
  \bibinfo{author}{\bibfnamefont{J.}~\bibnamefont{Schlipf}},
  \bibinfo{author}{\bibfnamefont{K.}~\bibnamefont{Held}}, \bibnamefont{and}
  \bibinfo{author}{\bibfnamefont{D.}~\bibnamefont{Vollhardt}},
  \bibinfo{journal}{Phys. Rev. B} \textbf{\bibinfo{volume}{58}},
  \bibinfo{pages}{12749 } (\bibinfo{year}{1998}).
  
\bibitem[{\citenamefont{Arrachea}(2000)}]{Arrachea:00}
\bibinfo{author}{\bibfnamefont{L.}~\bibnamefont{Arrachea}},
  \bibinfo{journal}{Phys. Rev. B} \textbf{\bibinfo{volume}{62}},
  \bibinfo{pages}{10033 } (\bibinfo{year}{2000}).

\bibitem[{\citenamefont{Ulmke}(1998)}]{Ulmke:98}
\bibinfo{author}{\bibfnamefont{M.}~\bibnamefont{Ulmke}}, \bibinfo{journal}{Eur.
  Phys. J. B} \textbf{\bibinfo{volume}{1}}, \bibinfo{pages}{301 }
  (\bibinfo{year}{1998}).

\bibitem[{\citenamefont{Wegner et~al.}(1998)\citenamefont{Wegner, Potthoff, and
  Nolting}}]{Wegner:98}
\bibinfo{author}{\bibfnamefont{T.}~\bibnamefont{Wegner}},
  \bibinfo{author}{\bibfnamefont{M.}~\bibnamefont{Potthoff}}, \bibnamefont{and}
  \bibinfo{author}{\bibfnamefont{W.}~\bibnamefont{Nolting}},
  \bibinfo{journal}{Phys. Rev. B} \textbf{\bibinfo{volume}{57}},
  \bibinfo{pages}{6211 } (\bibinfo{year}{1998}).

\bibitem[{\citenamefont{Hlubina et~al.}(1997)\citenamefont{Hlubina, Sorella,
  and Guinea}}]{Hlubina:97}
\bibinfo{author}{\bibfnamefont{R.}~\bibnamefont{Hlubina}},
  \bibinfo{author}{\bibfnamefont{S.}~\bibnamefont{Sorella}}, \bibnamefont{and}
  \bibinfo{author}{\bibfnamefont{F.}~\bibnamefont{Guinea}},
  \bibinfo{journal}{Phys. Rev. Lett.} \textbf{\bibinfo{volume}{78}},
  \bibinfo{pages}{1343 } (\bibinfo{year}{1997}).

\bibitem[{\citenamefont{Obermeier et~al.}(1997)\citenamefont{Obermeier,
  Pruschke, and Keller}}]{Obermeier:97}
\bibinfo{author}{\bibfnamefont{T.}~\bibnamefont{Obermeier}},
  \bibinfo{author}{\bibfnamefont{T.}~\bibnamefont{Pruschke}}, \bibnamefont{and}
  \bibinfo{author}{\bibfnamefont{J.}~\bibnamefont{Keller}},
  \bibinfo{journal}{Phys. Rev. B}
  \textbf{\bibinfo{volume}{56}}, \bibinfo{pages}{R8479 } (\bibinfo{year}{1997}).

\bibitem[{\citenamefont{Zitzler et~al.}(2002)\citenamefont{Zitzler, Pruschke,
  and Bulla}}]{Zitzler:02}
\bibinfo{author}{\bibfnamefont{R.}~\bibnamefont{Zitzler}},
  \bibinfo{author}{\bibfnamefont{T.}~\bibnamefont{Pruschke}}, \bibnamefont{and}
  \bibinfo{author}{\bibfnamefont{R.}~\bibnamefont{Bulla}},
  \bibinfo{journal}{Eur. Phys. J. B} \textbf{\bibinfo{volume}{27}},
  \bibinfo{pages}{473 } (\bibinfo{year}{2002}).

\bibitem[{\citenamefont{Werner et~al.}(2006)\citenamefont{Werner, Comanac,
  Medici, Troyer, and Millis}}]{Werner:06}
\bibinfo{author}{\bibfnamefont{P.}~\bibnamefont{Werner}},
  \bibinfo{author}{\bibfnamefont{A.}~\bibnamefont{Comanac}},
  \bibinfo{author}{\bibfnamefont{L.}~\bibnamefont{de'Medici}},
  \bibinfo{author}{\bibfnamefont{M.}~\bibnamefont{Troyer}}, \bibnamefont{and}
  \bibinfo{author}{\bibfnamefont{A.J.}~\bibnamefont{Millis}},
  \bibinfo{journal}{Phys. Rev. Lett.} \textbf{\bibinfo{volume}{97}},
  \bibinfo{pages}{076405 } (\bibinfo{year}{2006}).

\bibitem[{\citenamefont{Haule}(2006)}]{Haule:06}
\bibinfo{author}{\bibfnamefont{K.}~\bibnamefont{Haule}},
  \bibinfo{journal}{Phys. Rev. B} \textbf{\bibinfo{volume}{75}},
  \bibinfo{pages}{155113 } (\bibinfo{year}{2007}).

\bibitem[{\citenamefont{Jarrell and Gubernatis}(1996)}]{Jarrell:96}
\bibinfo{author}{\bibfnamefont{M.}~\bibnamefont{Jarrell}} \bibnamefont{and}
  \bibinfo{author}{\bibfnamefont{J.}~\bibnamefont{Gubernatis}},
  \bibinfo{journal}{Phys. Rep.} \textbf{\bibinfo{volume}{269}},
  \bibinfo{pages}{133 } (\bibinfo{year}{1996}).

\bibitem[{\citenamefont{Kajoeter et~al.}(1996)\citenamefont{Kajoeter, Kotliar,
  and Moeller}}]{Kajoeter:96}
\bibinfo{author}{\bibfnamefont{H.}~\bibnamefont{Kajueter}},
  \bibinfo{author}{\bibfnamefont{G.}~\bibnamefont{Kotliar}}, \bibnamefont{and}
  \bibinfo{author}{\bibfnamefont{G.}~\bibnamefont{Moeller}},
  \bibinfo{journal}{Phys. Rev. B} \textbf{\bibinfo{volume}{53}},
  \bibinfo{pages}{16214 } (\bibinfo{year}{1996}).

\bibitem[{\citenamefont{Polyakov, A.M.}(1975)}]{Polyakov:75}
\bibinfo{author}{\bibfnamefont{A.M.}~\bibnamefont{Polyakov}},
  \bibinfo{journal}{Phys. Lett. B} \textbf{\bibinfo{volume}{59B}},
  \bibinfo{pages}{79 } (\bibinfo{year}{1975}).

\bibitem[{\citenamefont{Jaksch and Zoller}(2005)}]{Jaksch:05}
\bibinfo{author}{\bibfnamefont{D.}~\bibnamefont{Jaksch}} \bibnamefont{and}
  \bibinfo{author}{\bibfnamefont{P.}~\bibnamefont{Zoller}},
  \bibinfo{journal}{Ann. Phys.} \textbf{\bibinfo{volume}{315}},
  \bibinfo{pages}{52 } (\bibinfo{year}{2005}).

\bibitem[{\citenamefont{Georges}(2007)}]{Georges:07}
\bibinfo{author}{\bibfnamefont{A.}~\bibnamefont{Georges}},
  \bibinfo{journal}{cond-mat/0702122}  (\bibinfo{year}{2007}).

\end{thebibliography}

\end{document}